%% file: acl_latex.tex
\documentclass[11pt]{article}

\usepackage[preprint]{acl}

\usepackage{times}
\usepackage{latexsym}

\usepackage[T1]{fontenc}

\usepackage[utf8]{inputenc}

\usepackage{microtype}

\usepackage{inconsolata}

\usepackage{graphicx}

\usepackage[inkscapelatex=false]{svg}

\usepackage{booktabs} 
\usepackage{array}    
\usepackage{pifont}   
\usepackage{tabularx}
\usepackage{listings}
\usepackage{xcolor}
\usepackage{hyperref}
\usepackage{subcaption}
\usepackage{amsmath}
\usepackage{enumitem}
\usepackage{multirow}
\usepackage{xcolor}
\usepackage{colortbl}
\usepackage{tcolorbox}
\tcbuselibrary{breakable}
\usepackage{float}
\usepackage{cuted}
\usepackage{stfloats}
\usepackage{fontawesome5}

\newcommand{\tightparagraph}[1]{\vspace{-4pt} \paragraph{#1}}

\colorlet{punct}{red!60!black}
\definecolor{background}{HTML}{EEEEEE}
\definecolor{delim}{RGB}{20,105,176}
\colorlet{numb}{magenta!60!black}
\lstdefinelanguage{json}{
    basicstyle=\small\ttfamily,
    numberstyle=\scriptsize,
    stepnumber=1,
    numbersep=8pt,
    showstringspaces=false,
    breaklines=true,
    frame=lines,
    backgroundcolor=\color{background},
    literate=
     *{0}{{{\color{numb}0}}}{1}
      {1}{{{\color{numb}1}}}{1}
      {2}{{{\color{numb}2}}}{1}
      {3}{{{\color{numb}3}}}{1}
      {4}{{{\color{numb}4}}}{1}
      {5}{{{\color{numb}5}}}{1}
      {6}{{{\color{numb}6}}}{1}
      {7}{{{\color{numb}7}}}{1}
      {8}{{{\color{numb}8}}}{1}
      {9}{{{\color{numb}9}}}{1}
      {:}{{{\color{punct}{:}}}}{1}
      {,}{{{\color{punct}{,}}}}{1}
      {\{}{{{\color{delim}{\{}}}}{1}
      {\}}{{{\color{delim}{\}}}}}{1}
      {[}{{{\color{delim}{[}}}}{1}
      {]}{{{\color{delim}{]}}}}{1},
}

\title{ScaleBox: Enabling High-Fidelity and Scalable Code Verification for Large Language Models}

\author{
  Jiasheng Zheng${}^{1,2}$\thanks{Equal contribution.},
  Xin Zheng${}^{1}$\footnotemark[1],
  Boxi Cao${}^{1}$\footnotemark[1],
  \textbf{Pengbo Wang}${}^{1,2}$,
  \textbf{Zhengzhao Ma}${}^{1,2}$,
  \\
  \textbf{Qiming Zhu}${}^{1,2}$,
  \textbf{Jiazhen Jiang}${}^{1,2}$,
  \textbf{Yaojie Lu${}^{1}$}\thanks{Corresponding authors.},
  \textbf{Hongyu Lin${}^{1}$}\footnotemark[2],
  \textbf{Xianpei Han${}^{1}$},
  \textbf{Le Sun${}^{1}$}
  \\
  ${}^{1}$Chinese Information Processing Laboratory \\
  Institute of Software, Chinese Academy of Sciences \\
  ${}^{2}$University of Chinese Academy of Sciences \\
 {\tt \{zhengjiasheng2022,zhengxin2020,caoboxi,luyaojie,hongyu\}@iscas.ac.cn} \\
 \faGithub \text{ } \url{https://github.com/icip-cas/ScaleBox}
}

\begin{document}
\maketitle
\begin{abstract}
Code sandboxes have emerged as a critical infrastructure for advancing the coding capabilities of large language models, providing verifiable feedback for both RL training and evaluation. 
However, existing systems fail to provide accurate verification and efficiency under high-concurrency workloads.
We present \textsc{ScaleBox}, a high-fidelity and scalable system designed to address these limitations in large-scale code training.
\textsc{ScaleBox} introduces automated special-judge generation and management, fine-grained parallel execution across test cases with seamless multi-node coordination, and a configuration-driven evaluation suite for reproducible benchmarking.
A series of experiments demonstrates that \textsc{ScaleBox} significantly enhances code verification accuracy and efficiency.
Our further RLVR experiments show that \textsc{ScaleBox} substantially improves both performance on LiveCodeBench and training stability, significantly outperforming heuristic-matching baselines.
By providing a reliable and high-throughput infrastructure, \textsc{ScaleBox}\footnote{The demonstration video is available at: \url{https://youtu.be/TGW-qxRFb5s}.} facilitates more effective research and development in large-scale code training.

\end{abstract}

\section{Introduction}

\begin{figure}[h]
  \centering
  \includegraphics[width=0.48\textwidth]{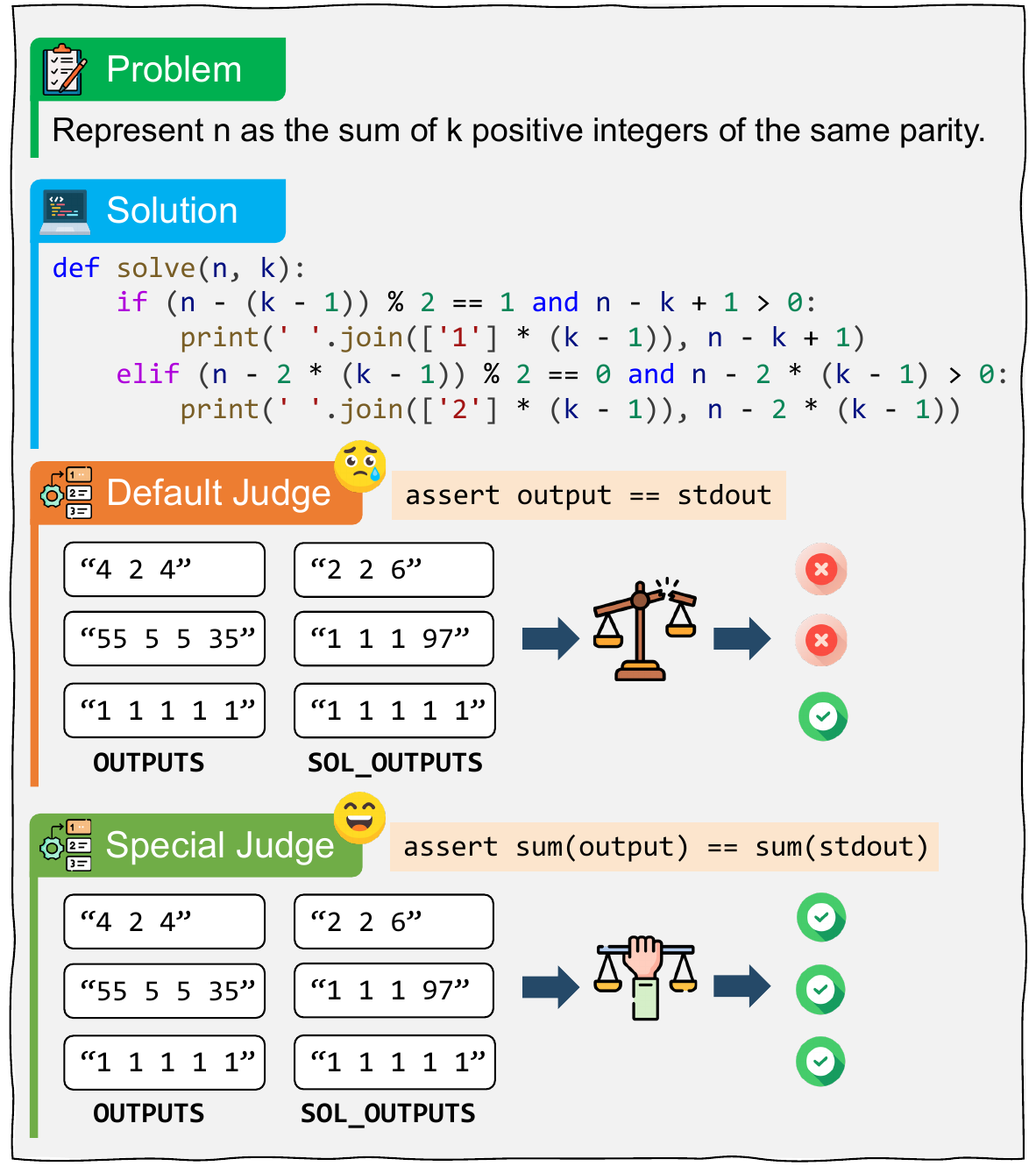}
  \caption{Comparison between Exact Match and Special Judge evaluation.}
  \label{tab:figure_head}
\end{figure}

Large Language Models (LLMs) have achieved substantial advances in code intelligence~\citep{zhang2024unifying,yang2025code}, reshaping code agents and automated software engineering~\citep{dong2025survey,wang2025ai,he2025llm,jiang2026survey}.
A central enabler of this progress is the code sandbox~\citep{guo2025deepseek,team2025kimik1_5,team2025kimik2}, an essential infrastructure that provides executable ground-truth feedback for both evaluation and Reinforcement Learning from Verifiable Rewards (RLVR).
In this paradigm, the accuracy and throughput of sandbox feedback have become key bottlenecks for evaluation reliability and training efficiency.

However, when scaling to large-scale code training, existing sandbox systems exhibit substantial deficiencies in both verification accuracy and efficiency.
First, current systems suffer from a \textbf{verification accuracy gap} due to their excessive reliance on exact match (EM) or heuristic-based matching evaluation~\citep{cheng2024fullstack,dou-etal-2025-multi,journals/tse/CassanoGNNPPYZAFGGJ23}.
Such mechanisms incorrectly assume a single canonical output, ignoring tasks with multiple valid solutions or precision requirements (Figure~\ref{tab:figure_head}).  
Our analysis of 34,757 problems reveals that 14.57\% of tasks require specialized verification (Special Judges), and 59.01\% of their correct solutions are falsely rejected by EM. 
This misclassification injects significant noise into learning signals, leading to biased gradients and suboptimal convergence.
Second, existing code verification pipelines suffer from a critical \textbf{efficiency bottleneck}. Most systems employ static deployment and coarse-grained task scheduling~\citep{cheng2024fullstack,feng2025retool}, hindering horizontal scaling in multi-node clusters.
This leads to a severe computational asymmetry: accelerators (e.g., GPUs and NPUs) are heavily utilized while host processors remain underused for parallel code execution and testing, leading to host–accelerator imbalance that limits throughput and scalability in large-scale training.

To this end, we present \textsc{ScaleBox}, a high-fidelity and scalable sandbox system engineered for large-scale code RLVR training and evaluation. 
First, \textsc{ScaleBox} introduces an automated synthesis and management mechanism for special judges, together with a unified verification framework. This design enables adaptive judgment logic, substantially fortifying the accuracy and stability of rewards. 
Second, \textsc{ScaleBox} implements hybrid parallelism at both the instance and unit-test levels, along with seamless multi-node deployment. Controlled via a centralized web-based dashboard, the system supports hot-update operations across distributed clusters, as well as real-time resource and log monitoring, thereby fully leveraging idle CPU resources for high-throughput execution.
Third, \textsc{ScaleBox} provides a configuration-driven evaluation suite supporting major code benchmarks, offering the research community a stable, efficient, and reproducible evaluation platform.

A series of experiments demonstrates that \textsc{ScaleBox} significantly enhances both verification accuracy and efficiency. 
Accuracy evaluation on AetherCode~\citep{wang2025aethercode} reveals that our synthetic special judges achieve over 84\% per-solution precision, indicating high-fidelity verification.
Efficiency evaluation shows that \textsc{ScaleBox} achieves a 59\% speedup over verl’s native execution and exhibits substantial throughput scalability across multi-node clusters.
To further assess its practical impact, we conduct RLVR training using \textsc{ScaleBox} as the verification backbone. 
Experimental results show that the high-fidelity feedback provided by \textsc{ScaleBox} substantially reduces reward noise, leading to significant improvements in training performance and stability. 
Specifically, Qwen3-8B~\citep{yang2025qwen3} trained with \textsc{ScaleBox} achieves substantial performance gains on the widely used LCB-V5 and LCB-V6 benchmarks~\citep{jain2025livecodebench}, significantly outperforming heuristic-matching baselines.

Our main contributions are summarized as:
\begin{itemize}
    \item We propose \textsc{ScaleBox}, a high-fidelity and scalable system designed to enhance verification accuracy and efficiency for large-scale code training and evaluation.
    \item We develop a configuration-driven evaluation framework supporting major code benchmarks to streamline assessment workflows.
    \item Experiments show that \textsc{ScaleBox} significantly improves verification accuracy and efficiency, leading to better performance and stability in RLVR training.
\end{itemize}

\section{Related Works}

\subsection{Reinforcement Learning with Verifiable Rewards}

Reinforcement Learning from Verifiable Rewards (RLVR) has emerged as a powerful paradigm for enhancing LLM reasoning via objective feedback~\citep{guo2025deepseek}. 
While discussions continue regarding whether RL elicits latent behaviors or expands intrinsic capabilities~\citep{yue2025does}, recent empirical evidence suggests that RLVR can synthesize novel strategies surpassing distillation-based limits~\citep{wen2025reinforcement,liu2025prorl,chen2025acereason}.
To ensure stable policy scaling and prevent entropy collapse~\citep{cui2025entropy}, algorithms like DAPO~\citep{yu2025dapo}, Dr. GRPO~\citep{liu2025understanding}, and GSPO~\citep{zheng2025group} introduce diversity-preserving and bias-correction mechanisms.
However, the efficacy of these methods is strictly bounded by the fidelity and scalability of the verification signal~\citep{liu-etal-2025-compassverifier, chen2025xverify}.
Although tool-augmented~\citep{feng2025cosineverifier} and reasoning-augmented approaches~\citep{zheng2025sci} have been proposed, providing high-fidelity judgment for RL training remains a critical frontier.

\subsection{Code Execution Environment}

The emergence of RLVR in the code domain demands execution environments that provide both high-throughput scaling and high-precision reward generation. 
However, existing systems remain fragmented.
MultiPL-E~\citep{journals/tse/CassanoGNNPPYZAFGGJ23} pioneered large-scale multi-language evaluation, but its offline execution is unsuitable for the iterative feedback loops required in RL. 
While recent works like SandboxFusion~\citep{cheng2024fullstack} and MPLSandbox~\citep{dou-etal-2025-multi} introduce RL integration for training, they lack either the distributed infrastructure for high-throughput scaling or the specialized judging logic necessary for generating the precise, verifiable rewards that RLVR depends on. 
Conversely, systems like Judge~\citep{fu2025klear} prioritize evaluation accuracy through special judges but lack RL compatibility and scalability. 
ScaleBox fills this critical gap by providing a scalable sandbox system that simultaneously optimizes for efficiency and accuracy to support the full lifecycle of RLVR.

\begin{figure*}[t]
  \centering
  \includegraphics[width=\textwidth]{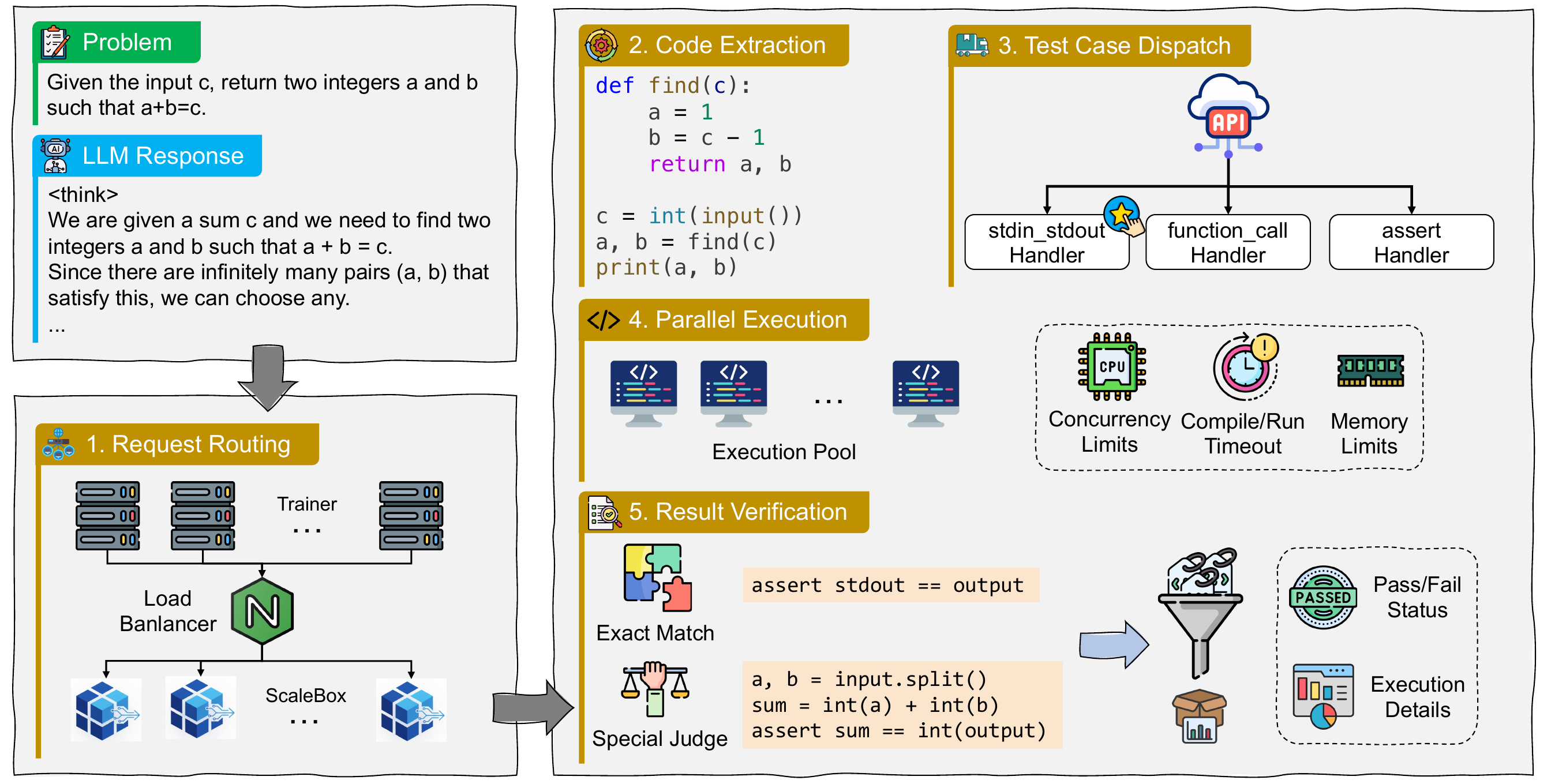}
  \caption{Overview of the \textsc{ScaleBox} Architecture. A distributed architecture featuring NGINX load balancing, test-case parallelism, and unified verification with special judge support, optimized for high-throughput and reproducible code training and evaluation.}
  \label{fig:scalebox_architecture}
\end{figure*}

\section{Preliminary Study: The Fragility of Exact Match Evaluation}

Current code execution frameworks predominantly rely on Exact Match (EM) of string outputs to verify program correctness \citep{dou-etal-2025-multi, cheng2024fullstack}. 
While computationally efficient, EM fails to capture the semantic correctness of tasks where multiple valid representations exist. 
For instance, problems involving multiple valid outputs (e.g., ``output any valid sequence'') or floating-point arithmetic (with precision tolerances) are inherently incompatible with string-based evaluation.

To quantify this misalignment, we perform a systematic study of the PrimeIntellect~\citep{2025synthetic1} dataset, covering 34,757 Python problems. 
We use \textit{Qwen3-235B}~\citep{yang2025qwen3} to determine whether each problem requires special judge evaluation beyond exact string matching (see Figure~\ref{tab:special_judge_classification_prompt}).
\textbf{Our analysis demonstrates that EM verification is fundamentally ill-suited for a significant portion of code tasks, leading to systematic misclassification of correct solutions.}
Specifically, we find that 14.57\% of the problems are inherently incompatible with EM evaluation, which requires a special judge to verify correctness, a feature currently absent from most mainstream RL training sandboxes.
Among them, 78.26\% admit multiple valid outputs and 21.74\% require floating-point comparisons with specified numerical tolerances.
More critically, we observe that 59.01\% of the ground-truth solutions fail EM-based unit tests, despite being logically correct.

These findings indicate that naive string-based evaluation would systematically reject a large fraction of correct solutions for problems requiring special judges. 
In RLVR, such false negatives introduce noisy reward signals, biasing the learning process and undermining training stability. 
This motivates the need for a sandbox system with native support for special judge evaluation to ensure accurate verification across diverse programming problem types.

\section{ScaleBox System Architecture}
\label{sec:scalebox_architecture}

\textsc{ScaleBox} is a distributed, high-concurrency sandbox infrastructure designed for secure code execution and evaluation. 
Built upon the foundation of SandboxFusion~\citep{cheng2024fullstack}, \textsc{ScaleBox} is specifically optimized to meet the rigorous demands of RL training and large-scale code benchmarking. 
As illustrated in Figure~\ref{fig:scalebox_architecture}, the architecture comprises three synergistic components: 
(i) \textbf{Evaluation Workflow} (\S\ref{sec:evaluation_workflow}): An optimized end-to-end pipeline for code extraction, execution, and verification.
(ii) \textbf{Distributed Deployment} (\S\ref{sec:distributed_deployment}): A scalable architecture designed for high-throughput RL training.
(iii) \textbf{Special Judge Evaluation} (\S\ref{sec:special_judge_evaluation}): A flexible verification framework for non-deterministic programming tasks.

\subsection{Evaluation Workflow}
\label{sec:evaluation_workflow}

The \textsc{ScaleBox} evaluation pipeline is designed to minimize latency while ensuring execution fidelity. 
The workflow follows a five-stage process:

\begin{enumerate}[leftmargin=16pt, itemsep=0pt, topsep=4pt]
\item \textbf{Request Routing}: Incoming evaluation tasks are managed by an NGINX-based load balancer, which dynamically routes requests to available sandbox workers. 
\item \textbf{Code Extraction}: The worker node uses robust heuristics to extract executable code from diverse LLM formats, including Markdown blocks, code fragments, and scripts.
\item \textbf{Test Case Dispatch}: Utilizing the unified API, the system identifies the test type (\texttt{stdin\_stdout}, \texttt{function\_call}, or \texttt{assert}) and generates language-specific test harnesses. 
\item \textbf{Parallel Execution}: Test cases are executed in parallel within isolated environments, with configurable constraints on memory, CPU usage, and multi-level wall-clock time. The system maps time constraints to different lifecycle stages, covering individual compilation, per-test execution, and global session timeouts.
\item \textbf{Multi-Stage Verification}: Prioritizes optimized \textit{exact match} (handling whitespace and float tolerance), falling back to a \textit{special judge} for custom verification if required.
\end{enumerate}

To facilitate standardized RLVR research and rapid iteration, \textsc{ScaleBox} further
provides a one-click evaluation workflow for widely used code benchmarks, such as LiveCodeBench~\citep{jain2025livecodebench}, HumanEval~\citep{chen2021evaluating}, and AetherCode~\citep{wang2025aethercode}.

\subsection{Distributed Deployment}
\label{sec:distributed_deployment}

To accommodate the bursty and high-volume workloads characteristic of code RL, \textsc{ScaleBox} implements a horizontally scalable architecture.

\tightparagraph{Load Balancing and Availability} 
An NGINX-based ingress controller implements a round-robin strategy to distribute workloads. 
To ensure robustness during long-running training jobs, we utilize Docker-based health checks and automated recovery protocols, maintaining high availability even under extreme hardware utilization.

\tightparagraph{Hybrid Parallelism} 
Unlike existing systems that parallelize only at the instance level, \textsc{ScaleBox} introduces \textit{hybrid parallelism} at both the instance and unit-test levels.
Each worker node supports configurable multi-tenancy, processing multiple instances concurrently while executing individual test cases within a single instance in parallel.

\tightparagraph{Web-based Dashboard}
\textsc{ScaleBox} provides a web-based dashboard (Appendix~\ref{sec:dashboard}) for visualized distributed deployment, resource monitoring, and log inspection. 
Through this interface, we can perform hot updates across multi-node clusters, dynamically reallocating and utilizing idle CPU resources to achieve efficient and scalable execution.

\tightparagraph{Efficiency Benchmarking} 

\begin{table}[t]
\centering
\small
\resizebox{\columnwidth}{!}{
\begin{tabular}{l >{\centering\arraybackslash}p{0.7cm} cc}
\toprule
\textbf{Method} & \textbf{Nodes} & \textbf{Time (s)} & \textbf{Throuput (tasks/s)} \\
\midrule
verl \tiny{Prime} & 1 & 331.25 & 24.73 (\scalebox{0.9}{1.00$\times$}) \\
SandboxFusion & 1 & 548.93 & 14.92 (\scalebox{0.9}{0.56$\times$}) \\
\midrule
\textsc{ScaleBox} & 1 & 208.38 & 39.31 (\scalebox{0.9}{1.59$\times$}) \\
\textsc{ScaleBox} & 2 & 163.40 & 50.13 (\scalebox{0.9}{2.03$\times$}) \\
\textsc{ScaleBox} & 3 & 131.92 & 62.10 (\scalebox{0.9}{2.51$\times$}) \\
\bottomrule
\end{tabular}
}
\caption{Efficiency comparison of \textsc{ScaleBox} against widely-adopted strong baselines.}
\label{tab:efficiency}
\end{table}

We evaluate \textsc{ScaleBox} using 8,192 Python problems from the PrimeIntellect~\citep{2025synthetic1} dataset on Intel Xeon 6700P Series CPUs (64 cores per node). 
Our results (Table~\ref{tab:efficiency}) show that \textsc{ScaleBox} achieves a single-node throughput of 39.31 tasks/s, representing a 1.59$\times$ improvement over verl~\citep{sheng2025hybridflow} framework (24.73 tasks/s) with Prime~\citep{cui2025process} code and 2.63$\times$ over SandboxFusion~\citep{cheng2024fullstack} baseline (14.92 tasks/s).
This throughput enhancement reflects the architectural advantages of \textsc{ScaleBox}, particularly its integration of hybrid parallelism and batch processing, which are specifically engineered to mitigate the serial execution bottlenecks prevalent in existing sandboxes.
Furthermore, \textsc{ScaleBox} scales to 62.10 tasks/s on 3 nodes, demonstrating substantial scalability for large-scale RL clusters.

\subsection{Special Judge Evaluation}
\label{sec:special_judge_evaluation}

Standard exact-match verification often yields false negatives in programming tasks with multiple valid solutions, such as those involving non-deterministic algorithms, multiple optimal paths, or floating-point precision~\citep{chou2025autocodebench,wang2025aethercode}. 
\textsc{ScaleBox} addresses this via a generalized special judge framework.

For stdin/stdout tasks, the special judge acts as a programmable verifier that reads three files: (1) \texttt{stdin.txt} (the input), (2) \texttt{stdout.txt} (the reference output), and (3) \texttt{answer.txt} (the participant output). 
The judge program returns a binary verdict based on task-specific logic.

To optimize performance, \textsc{ScaleBox} employs a \textbf{Short-Circuit Verification} strategy: the computationally expensive special judge is only invoked if the participant output fails the initial exact-match heuristic. This two-stage approach ensures both rigorous correctness and high execution efficiency.

\section{Automated Special Judge Synthesis}

In this section, we present an automated pipeline to synthesize \textit{special judges}, mitigating reward noise from exact matching.
We first detail the synthesis process and then validate judge fidelity to establish a reliable foundation for downstream RL.

\subsection{Synthesis Pipeline}
\label{sec:special_judge_synthesis}

\textsc{ScaleBox} transforms rigid exact-match signals into functional verification logic through a three-stage pipeline: (1) problem taxonomy detection, (2) program synthesis, and (3) sandbox-based verification. The implementation details can be found in Appendix~\ref{sec:special_judge_synthesis_details}.

\tightparagraph{Problem Taxonomy Detection}

We use an LLM to classify tasks based on their problem descriptions. The LLM identifies tasks where exact-match is insufficient and maps them into a predefined taxonomy:
(1) \textbf{non-deterministic outputs}, which allow multiple valid trajectories (e.g., any valid path in a DAG), 
or (2) \textbf{numerical tolerance}, which require precision-based comparisons within a specific threshold $\epsilon$ (e.g., ``absolute error $< 10^{-6}$'').

\tightparagraph{Program Synthesis}

For identified problems, we use an LLM to synthesize a Python-based judge program \texttt{validate\_solution}.
This judge operates by reading the problem input (\texttt{stdin.txt}), reference output (\texttt{stdout.txt}), and participant output (\texttt{answer.txt}) to produce a final binary verdict based on problem-specific constraints.

\tightparagraph{Sandbox-based Verification Loop}

To ensure reliability, synthesized judges undergo a pre-deployment validation stage within the \textsc{ScaleBox} environment. 
This involves a fidelity test to ensure the judge accepts ground-truth solutions and a robustness test to ensure it rejects null or known-incorrect outputs. 
This dual-verification filter discards malformed judges and triggers iterative LLM-based regeneration upon failure.

\subsection{Evaluation of Judge Fidelity}
\label{sec:special_judge_fidelity}

To assess the reliability of synthesized special judges, we conduct a systematic evaluation using 27 complex problems from the AetherCode dataset~\citep{wang2025aethercode} that require non-trivial verification logic. 
For each problem, we sample special judge codes up to 20 attempts. We then evaluate the valid sampled special judges against human-authored gold-standard oracles on a total of 269 correct and 260 incorrect submissions, focusing on two critical metrics:
\begin{itemize}[leftmargin=16pt, itemsep=0pt, topsep=4pt]
    \item True Positive Rate (TPR): The proportion of correct solutions that are correctly accepted by the generated judge.
    \item True Negative Rate (TNR): The proportion of incorrect solutions that are correctly rejected by the generated judge.
\end{itemize}

\begin{table}[t]
\centering
\resizebox{\columnwidth}{!}{
\begin{tabular}{lcc}
\toprule
\textbf{Model} & \textbf{TPR (\%)} & \textbf{TNR (\%)} \\
\midrule
GPT-5.2~\citep{singh2025openai} & 96.3 & 86.5  \\
Claude-3.7-Sonnet~\citep{claude_3_7} & 96.3 & 88.5  \\
DeepSeek-V3.2~\citep{liu2025deepseek} & 90.0 & 86.5  \\
Qwen3-235B~\citep{yang2025qwen3} & 90.3 & 84.2  \\
\bottomrule
\end{tabular}
}
\caption{Fidelity evaluation of synthesized special judges on 27 AetherCode instances. TPR and TNR are averaged across all submissions.}
\label{tab:special_judge_fidelity}
\end{table}

As shown in Table~\ref{tab:special_judge_fidelity}, \textbf{our synthesis method consistently generates highly effective verification logic across all evaluated LLMs}. 
Every model achieves a $\text{TPR} \ge 90.0\%$ and a $\text{TNR} \ge 84.0\%$, with Claude-3.7-Sonnet reaching the highest sensitivity (96.3\% TPR) and providing the strongest filtering (88.5\% TNR). 
These results demonstrate that synthesized judges can reliably distinguish between correct and incorrect code even for elite competitive programming tasks, validating the efficiency and high fidelity of our approach.

\section{RL Experiments}

Building on \textsc{ScaleBox}, we leverage RL to demonstrate that high-fidelity synthesized judges drive sustained gains by accurately verifying syntactically diverse programs where exact matching fails.

\begin{figure*}[t]
\centering
\begin{subfigure}[t]{0.48\textwidth}
\centering
\includegraphics[width=\textwidth]{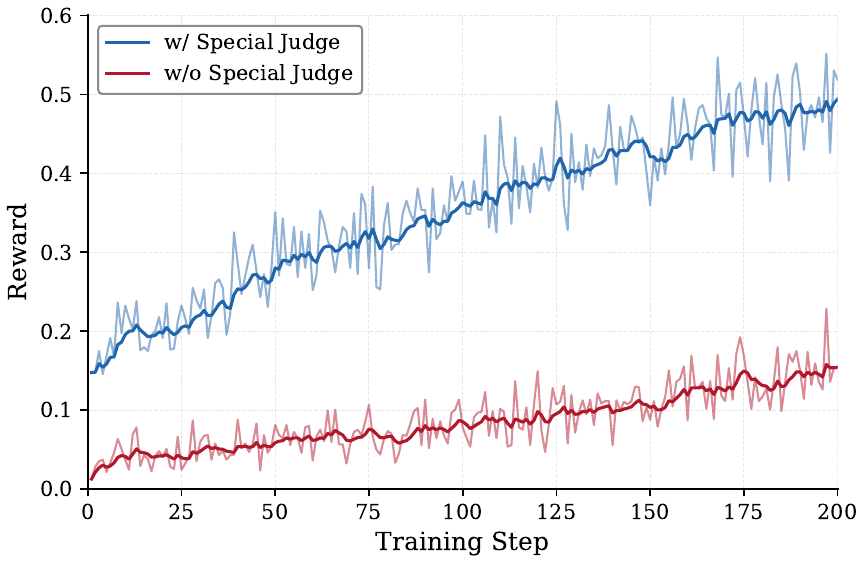}
\caption{Training reward trajectories.}
\label{fig:training_rewards}
\end{subfigure}
\hfill
\begin{subfigure}[t]{0.48\textwidth}
\centering
\includegraphics[width=\textwidth]{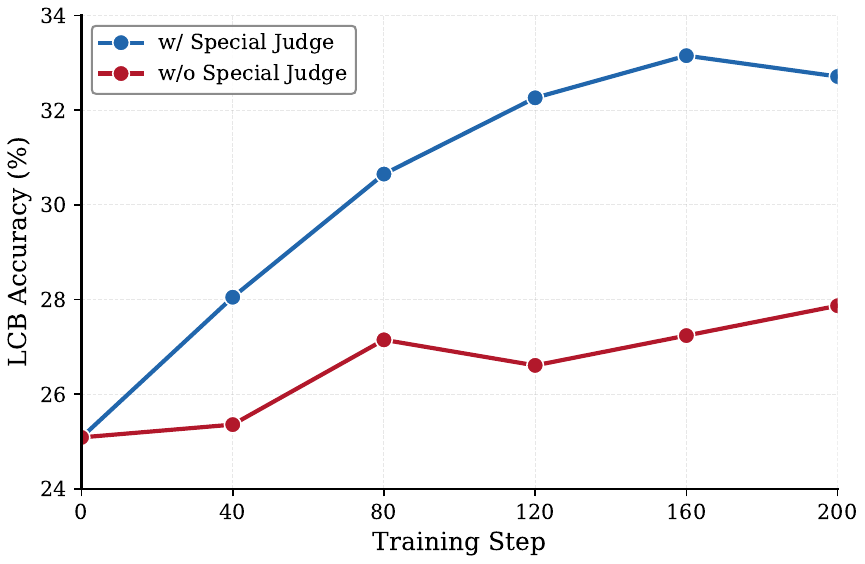}
\caption{Pass@1 Accuracy (LCB-v5).}
\label{fig:training_lcb_accuracy}
\end{subfigure}
\caption{Impact of reward fidelity on RL training (1.2K Subset): (a) Standard exact-match rewards are artificially suppressed due to false negatives, whereas Special Judges (SPJ) provide a more accurate and dense signal. (b) This higher-fidelity reward translates directly into superior and more stable Pass@1 performance across training steps.}
\vspace{-4pt}
\label{fig:training_progress}
\end{figure*}

\subsection{Experimental Setup}

Our training data consists of 26K Python problems filtered from the PrimeIntellect~\citep{2025synthetic1} ($\ge 5$ unit tests).
Within this set, our pipeline (\S\ref{sec:special_judge_synthesis}) using \textit{DeepSeek-V3.2}~\citep{liu2025deepseek} identifies 2.8K problems requiring Special Judges (SPJ). 
Notably, for 1.2K of these, reference solutions failed standard exact-match tests but are validated by our synthesized judges.
For RL setup, based on \textit{Qwen3-8B} (Non-thinking)~\citep{yang2025qwen3}, we perform the GRPO~\cite{shao2024deepseekmath} algorithm using the verl~\citep{sheng2025hybridflow} framework. Further details are provided in Appendix~\ref{sec:rl_training_details}.

To investigate the impact of special judges, we evaluate training performance across two scales: 
(1) the 1.2K SPJ subset where special judges are critical, 
and (2) the full 26K dataset which mixes standard and 2.8K SPJ problems. 
Performance is measured by Pass@1 on LiveCodeBench~\citep{jain2025livecodebench} (v5 and v6 subsets). 

\begin{table}[t]
\centering
\resizebox{\columnwidth}{!}{
\begin{tabular}{llcc}
\toprule
\textbf{Training Set} & \textbf{Reward Type} & \textbf{LCB-v5} & \textbf{LCB-v6} \\ \midrule
(Base Model) & -- & 25.09 & 27.21 \\ \midrule
1.2K Subset & w/o SPJ (EM) & 27.24 & 27.94 \\
\rowcolor{blue!5}
 & \textbf{w/ SPJ} (Ours) & \textbf{33.15} & \textbf{32.35} \\ \midrule
26K Full Dataset & w/o SPJ (EM) & 37.19 & 34.12 \\
\rowcolor{blue!5}
 & \textbf{w/ SPJ} (Ours) & \textbf{38.17} & \textbf{36.03} \\ \bottomrule
\end{tabular}
}
\caption{Comparative performance of Pass@1 (\%) on LiveCodeBench using Qwen3-8B as the base policy. Results show that incorporating Special Judges (SPJ) consistently improves performance across both datasets, demonstrating the critical role of reward fidelity.}
\vspace{-4pt}
\label{tab:rl_performance}
\end{table}

\subsection{Results and Analysis}

Table~\ref{tab:rl_performance} presents the comparative performance of RL training across the 1.2K special judge subset and the full 26K dataset.

\textbf{Synthesized special judges significantly improve RL performance by providing high-fidelity reward signals.} 
On the 1.2K special judge subset, SPJ-enhanced training achieves +5.91\% on LCB-v5. 
As illustrated in Figure~\ref{fig:training_rewards}, this gain is directly attributable to escaping the exact-match trap, where valid programs were previously mislabeled as failures. 
By resolving these false negatives, our method ensures that the policy is accurately rewarded for legitimate exploration.

\textbf{High-fidelity rewards enhance training stability and convergence.} 
Figure~\ref{fig:training_lcb_accuracy} demonstrates that the SPJ-enhanced model consistently maintains a performance lead throughout the training steps.
This gap suggests that cleaner reward signals reduce variance in credit assignment, allowing the model to converge more efficiently on robust coding patterns.

\textbf{The benefits of special-judge verification generalize to large-scale datasets.} 
Even on the full 26K dataset, where SPJ cases account for only $\sim 10\%$ of the total, we observe +1.91\% on LCB-v6. 
This disproportionate impact suggests that accurate evaluation of complex, non-deterministic problems provides a high-quality training signal that positively transfers to the model's overall reasoning capabilities, even in standard coding tasks.

\section{Conclusion}

In this paper, we introduce \textsc{ScaleBox}, a robust and scalable sandbox system that enhances reward accuracy and verification efficiency in large-scale code training. 
By leveraging automated special judge synthesis, unified verification, and parallel multi-node execution, \textsc{ScaleBox} mitigates the reward noise and host–accelerator imbalance issues present in existing systems. 
Our experiments confirm that \textsc{ScaleBox}'s high-fidelity rewards and efficient infrastructure yield substantial gains in model performance and training stability, providing a robust foundation for scaling training in LLMs.

\section*{Limitations}

Although \textsc{ScaleBox} is architecturally designed with configurations to support high scalability, its current evaluation is primarily focused on specific model scales and benchmarks. 
In future work, we plan to conduct larger-scale and multilingual RL experiments to fully verify the effectiveness and robustness of \textsc{ScaleBox} in more diverse and extensive code training scenarios. 
Additionally, while the current work focuses on providing feedback for code training and evaluation, we will explore applying \textsc{ScaleBox} to broader code agent scenarios, such as automated software engineering and multi-turn autonomous problem-solving, to further demonstrate its utility as a foundational infrastructure.

\section*{Acknowledgements}

We sincerely thank the reviewers for their insightful comments and valuable suggestions. This work was supported by the National Key R\&D Program of China (2024YFC3308000), Beijing Natural Science Foundation (L243006), the Natural Science Foundation of China (No. 62306303, 62476265). The authors would like to thank Huawei Ascend Cloud Ecological Development Project for the support of Ascend 910 processors.

\bibliography{custom}

\input{appendix}

\end{document}

%% file: appendix.tex
\clearpage

\appendix

\section{Screenshots of ScaleBox Dashboard}
\label{sec:dashboard}

\begin{figure*}[!b]
\centering
\includegraphics[width=0.49 \textwidth]{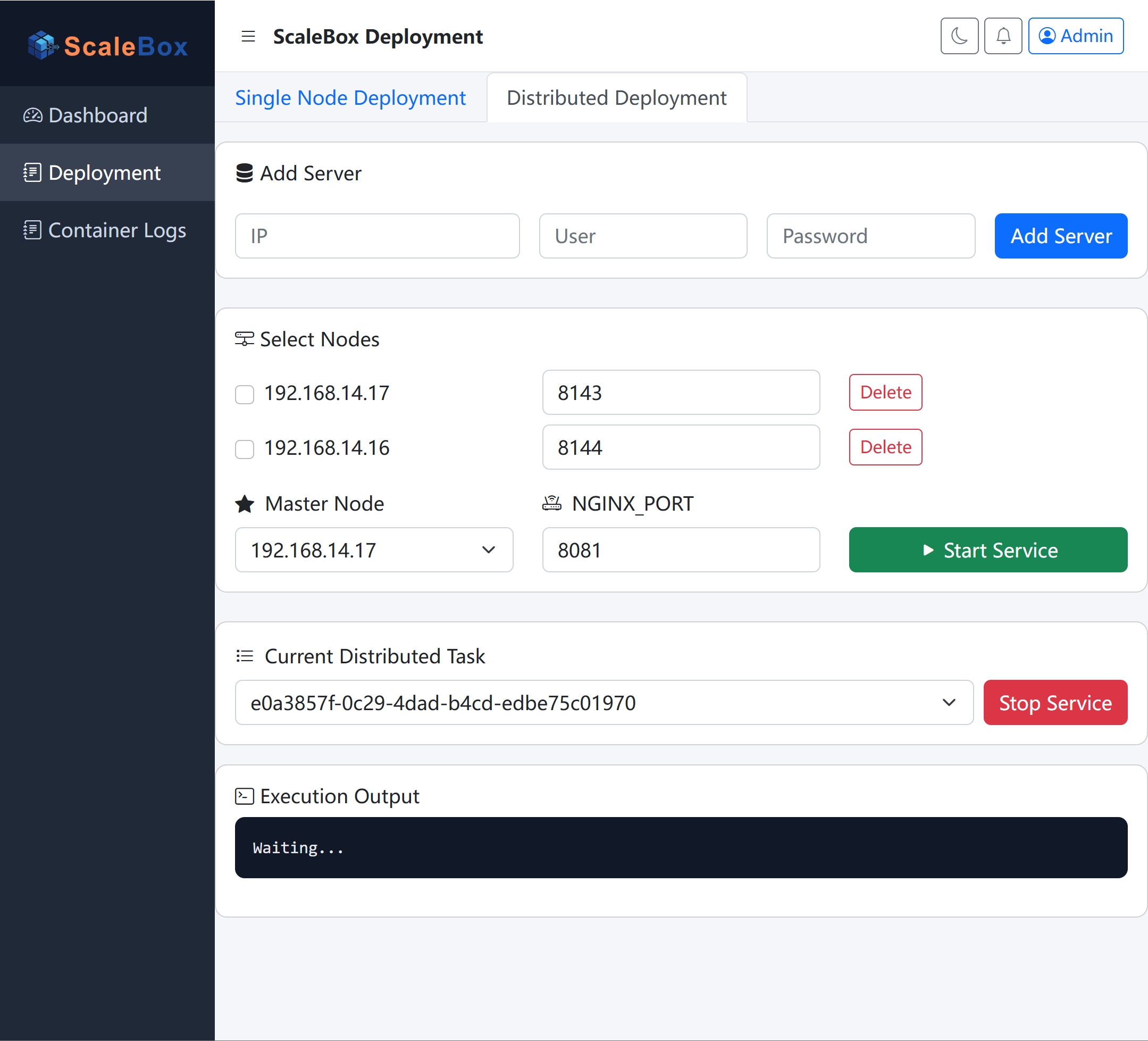}
\hfill
\includegraphics[width=0.49 \textwidth]{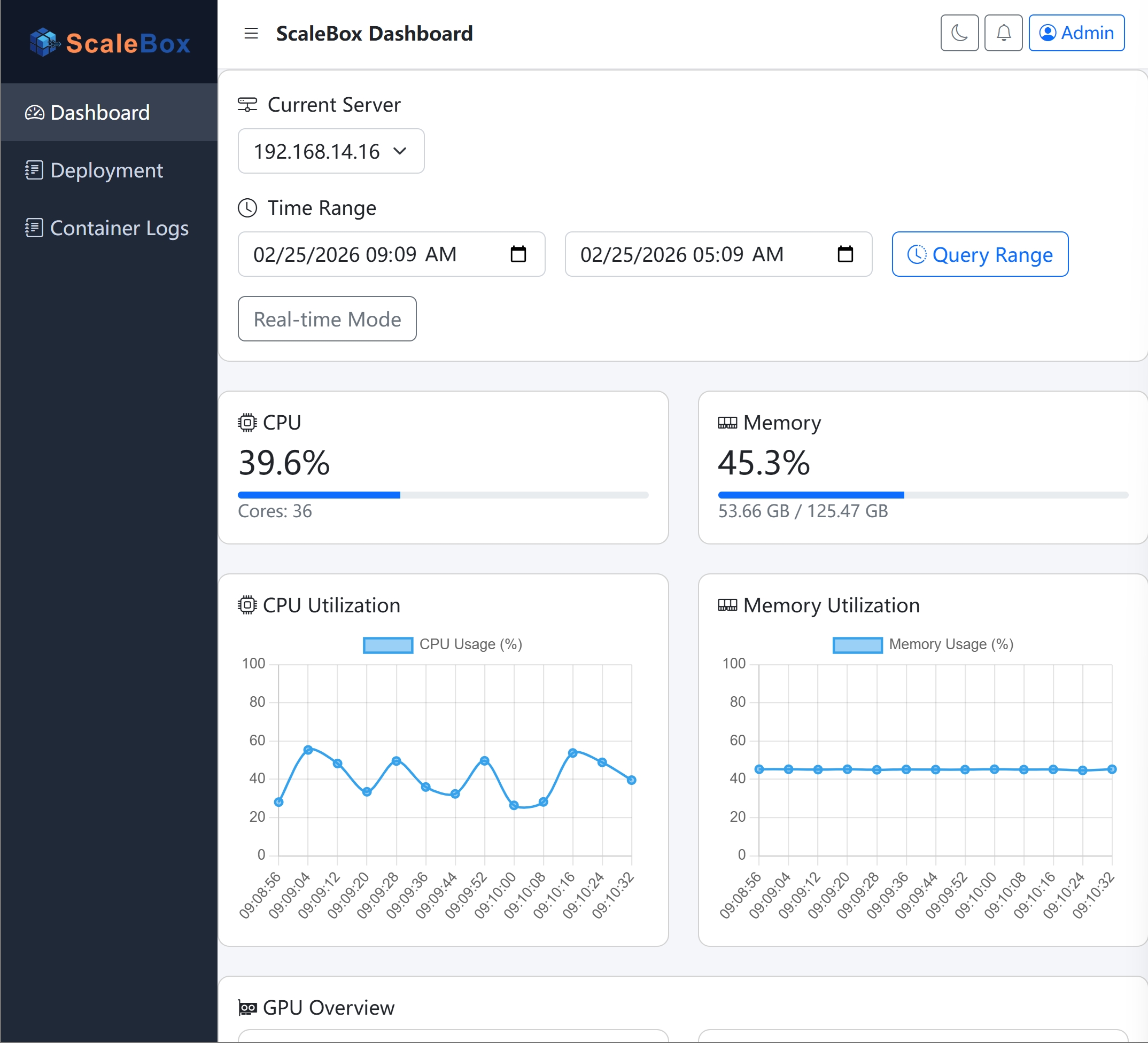}
\caption{
    The screenshots of the \textsc{ScaleBox} dashboard.
    \textbf{Left:} Distributed Deployment Panel.
    \textbf{Right:} Resource Monitoring Panel.
}
\label{fig:dashboard}
\end{figure*}

Figure~\ref{fig:dashboard} shows the screenshots of \textsc{ScaleBox} dashboard, which includes features for distributed deployment, resource monitoring, and log monitoring.

\section{Supported Benchmark of ScaleBox}

\begin{table*}[h]
\centering
\small
\resizebox{\textwidth}{!}{
\begin{tabular}{lcccccc}
\toprule
\textbf{Model} & \textbf{HumanEval} & \textbf{MBPP} & \textbf{HumanEval+} & \textbf{MBPP+} & \textbf{LiveCodeBench} & \textbf{AetherCode} \\
\midrule
Llama-3-8B-Instruct & 60.98 & 62.76 & 57.93 & 54.76 & 10.48 & 0.20 \\
Llama-3.1-8B-Instruct & 70.73 & 66.74 & 65.24 & 57.67 & 6.18 & 0.20 \\
DeepSeek-R1-Distill-Qwen-1.5B & 47.56 & 40.28 & 44.51 & 37.30 & 16.13 & 0.07 \\
Qwen3-4B  & 89.63 & 82.67 & 85.37 & 73.81 & 53.92 & 8.07 \\
Qwen3-8B  & 88.41 & 85.95 & 80.48 & 73.28 & 60.09 & 9.18 \\
\bottomrule
\end{tabular}
}
\caption{Pass@1 accuracy (\%) on HumanEval, MBPP, HumanEval+, MBPP+, LiveCodeBench, and AetherCode.}
\label{tab:benchmark_eval_results}
\end{table*}

\textsc{ScaleBox} supports diverse code benchmarks, including:
\begin{itemize}[leftmargin=16pt, itemsep=0pt, topsep=4pt]
    \item \textbf{Assert-based}: HumanEval~\citep{chen2021evaluating}, MBPP~\citep{austin2021program}, HumanEval+, and MBPP+~\citep{liu2023your}
    \item \textbf{Multi-language}: MultiPL-E \citep{journals/tse/CassanoGNNPPYZAFGGJ23}
    \item \textbf{Standard I/O and function call}: LiveCodeBench~\citep{jain2025livecodebench},
    \item \textbf{Special Judge}: Aethercode~\citep{wang2025aethercode}
\end{itemize}

Table~\ref{tab:benchmark_eval_results} presents evaluation results on standard benchmarks using the one-click evaluation workflow of \textsc{ScaleBox}.

\section{Training Details of RLVR}
\label{sec:rl_training_details}

For training data, we use PrimeIntellect/verifiable-coding-problems~\citep{2025synthetic1} dataset as source, which curated from APPS~\citep{hendrycks2021measuring}, CodeContests~\citep{doi:10.1126/science.abq1158}, Codeforces~\citep{jur1cek2023codeforces}, and TACO~\citep{li2023taco}.
We then select 26K Python problems that contain non-empty reference solutions and $\ge 5$ test cases.

For RL setup, based on \textit{Qwen3-8B} (Non-thinking)~\citep{yang2025qwen3}, we perform the GRPO~\cite{shao2024deepseekmath} algorithm using the verl~\citep{sheng2025hybridflow} framework, with a global batch size of 128, a mini batch size of 32, 8 rollouts per question, learning rate of 1e-6, max response length of 8192, $\epsilon_{low}$ of 0.2 and $\epsilon_{high}$ of 0.28.

For evaluation, we use the widely adopted v5 (2408–2502) and v6 (2501–2504) subsets of LiveCodeBench~\citep{jain2025livecodebench}, consisting of 279 and 170 problems, respectively.
We sample 4 times per problem, with max output length of 32768, temperature of 0.6, and top\_p of 0.95.

\section{More Details on Special Judge Synthesis}
\label{sec:special_judge_synthesis_details}

The automatic special judge synthesis pipeline (\S\ref{sec:special_judge_synthesis}) uses two prompts. 
The classification prompt (Figure~\ref{tab:special_judge_classification_prompt}) detects problems requiring special judges by identifying (1) multiple valid outputs and (2) floating-point comparisons with tolerance constraints, and returns structured JSON with confidence scores for filtering.
The generation prompt (Figure~\ref{tab:special_judge_generation_prompt}) provides an example special judge program and instructs the LLM to produce a standardized program that reads \texttt{stdin.txt}, \texttt{stdout.txt}, and \texttt{answer.txt}, and outputs correctness via exit codes. This few-shot scheme yields reliable verification programs across problem types.

\begin{figure*}[t]
\begin{tcolorbox}[width=1\textwidth, fontupper=\small, colback=blue!2, boxrule=0.9pt] 
\begin{lstlisting}[basicstyle=\tiny\ttfamily, breaklines=true]
You are an assistant that classifies programming / algorithm / data processing tasks regarding SPECIAL JUDGE need.

Decide for the given task text:
1. multiple_solutions? For example:
 - If there are multiple answers, print any of them
 - If there are multiple solutions, you are allowed to print any of them.
 - If there are multiple possible solutions, print any of them.
 - If there are different possible orders with a correct answer, print any of them.
 - If there are multiple solutions, satisfying the problem condition(s), you can print any "one" solution.

2. float_comparison? (floating point answers, precision, tolerance, absolute/relative error, decimals)

Return JSON object:
{
 "reason": "<short justification <less than 160 words>",
 "needs_special_judge": <true|false>,
 "categories": [ zero or more of "multiple_solutions","float_comparison" ],
 "confidence": <float 0..1>
}

Rules:
 - needs_special_judge is true iff categories non-empty.
 - confidence: 0.9 clear indicators, 0.6 somewhat, 0.3 guess.
 - Keep output STRICT JSON. No extra keys.
\end{lstlisting}
\end{tcolorbox}

\caption{Prompt Template for classifying programming problems that require special judge support.}
\label{tab:special_judge_classification_prompt}
\end{figure*}

\begin{figure*}[t]
\begin{tcolorbox}[width=1\textwidth, fontupper=\small, colback=blue!2, boxrule=0.9pt] 
\begin{lstlisting}[basicstyle=\tiny\ttfamily, breaklines=true]
Here is an example of task description and a special judge program in Python.
<problem>
{PROBLEM}
</problem>

<special_judge_program>
{SPECIAL_JUDGE}
</special_judge_program>

Given the stdin, stdout, answer and programming task, write the JUDGE python program to check if the answer is as valid as the stdout. You may want to leverage stdout to save computation when needed.

```python
import sys

def read_file(filepath):
    with open(filepath, 'r') as f:
        return f.read().strip().split('\n')

def validate_solution(stdin_path, stdout_path, answer_path):
    stdin_lines = read_file(stdin_path)
    stdout_lines = read_file(stdout_path)
    participant_output = read_file(answer_path)

    if participant_output == [''] and stdout_lines != ['']:
        return False
    # ...

stdin_path = "stdin.txt"
stdout_path = "stdout.txt"
answer_path = "answer.txt"

is_valid = validate_solution(stdin_path, stdout_path, answer_path)

if is_valid:
    sys.exit(0)
else:
    sys.exit(1)
```
\end{lstlisting}
\end{tcolorbox}

\caption{Prompt Template for generating special judge programs.}
\label{tab:special_judge_generation_prompt}
\end{figure*}